\documentclass[aps,preprint,prd,showpacs,nofootinbib]{revtex4-1}
\pdfoutput=1
\usepackage[T1]{fontenc}
\usepackage[utf8]{inputenc}
\usepackage{graphicx}
\usepackage{bm}
\usepackage{times}
\usepackage{xcolor}

\usepackage{amsmath}
\usepackage{amssymb}
\usepackage{slashed}

\usepackage{natbib}
\usepackage{hyperref}

\graphicspath{{Figures/}}
\usepackage{multirow}
\usepackage{diagbox}
\usepackage{cancel}
\usepackage{subcaption} 
\usepackage{placeins} 

\colorlet{grey}{gray}

\usepackage{tikz}
\usetikzlibrary{decorations.pathmorphing,arrows.meta,positioning}

\usepackage{fancyhdr}
\usepackage{footmisc}

\newcommand{\be}{\begin{equation}}
\newcommand{\ee}{\end{equation}}
\newcommand{\ba}{\begin{eqnarray}}
\newcommand{\ea}{\end{eqnarray}}
\newcommand{\bs}{\begin{subequations}}
\newcommand{\es}{\end{subequations}}

\newcommand{\ie}{{\it i.e.}\ }

\newcommand{\grts}{\raise.3ex\hbox{$>$\kern-.75em\lower1ex\hbox{$\sim$}}}
\newcommand{\lets}{\raise.3ex\hbox{$<$\kern-.75em\lower1ex\hbox{$\sim$}}}

\newcommand{\bc}{\begin{center}}
\newcommand{\ec}{\end{center}}

\begin{document}
\title{Disappearing Tracks from Electroweak Vector Dark Matter at the LHC}
\author{Fabián Hernández-Pinto$^{1,2,3}$\footnote[1]{\href{mailto:fabianhernandezpinto@gmail.com}{\footnotesize fabianhernandezpinto@gmail.com}}}
\author{Patricio Escalona$^{4,5}$\footnote[2]{\href{mailto:patricioescalona96@gmail.com}{\footnotesize patricioescalona96@gmail.com}}}
\author{Alfonso Zerwekh$^{1,2,3}$\footnote[3]{\href{mailto:alfonso.zerwekh@usm.cl}{\footnotesize alfonso.zerwekh@usm.cl}}}

\affiliation{\footnotesize $^1$Departamento de F\'isica, Universidad T\'ecnica Federico Santa Mar\'ia, Avenida Espa\~na 1680, Valpara\'iso, Chile}

\affiliation{\footnotesize $^2$Millennium Institute for Subatomic Physics at High Energy Frontier --- SAPHIR, Fern\'andez Concha 700, Santiago, Chile}

\affiliation{\footnotesize $^3$Centro Cient\'ifico-Tecnol\'ogico de Valpara\'iso, Universidad T\'ecnica Federico Santa Mar\'ia, Avenida Espa\~na 1680, Valpara\'iso, Chile}

\affiliation{\footnotesize $^4$Departamento de F\'isica, Universidade Federal da Para\'iba, Caixa Postal 5008, 58051-970, Jo\~ao Pessoa, PB, Brazil}

\affiliation{\footnotesize $^5$International Institute of Physics, Universidade Federal do Rio Grande do Norte, Campus Universit\'ario, Lagoa Nova, Natal, RN, 59078-970, Brazil}

\vspace{0.4cm}
\begin{abstract}
We study the disappearing-track signature of a spin-1 minimal dark matter candidate transforming in the real five-dimensional representation of $SU(2)_L$ with vanishing hypercharge. Radiative electroweak corrections induce a compressed mass spectrum among the multiplet components, allowing the singly charged state $V^\pm$ to attain a macroscopic lifetime and decay via $V^\pm \to V^0 \pi^\pm$ inside tracking detectors. We recast the ATLAS search for disappearing tracks with the full Run-2 dataset of $137\text{ fb}^{-1}$, originally targeted at long-lived charginos, computing the tracklet reconstruction probability analytically from the published efficiency so that the full lifetime range is covered. The procedure reproduces the ATLAS limits for pure winos within $6\%$ in mass. The Run-2 data exclude quintuplet masses up to $1080\text{ GeV}$ for proper decay lengths $c\tau \simeq 250\text{ mm}$, and $M_V < 630\text{ GeV}$ for a mass splitting of $166\text{ MeV}$. For the radiative mass splitting predicted by the model with a small Higgs-portal coupling, masses between $120\text{ GeV}$ and $240\text{--}300\text{ GeV}$ are excluded. With $3\text{ ab}^{-1}$ and a Run-2-like analysis, the sensitivity would extend to $1750\text{ GeV}$ for vanishing background and to $1530\text{ GeV}$ for a linear extrapolation of the instrumental background.
\end{abstract}

\maketitle

\section{Introduction}\label{sec:intro}
The Standard Model (SM) provides an accurate description of particle interactions at accessible energies, but it does not account for several established phenomena, including neutrino masses and oscillations~\cite{Esteban:2024eli,JUNO:2025gmd,JUNO:2026tap}, the baryon asymmetry of the Universe~\cite{Bodeker:2020ghk,Planck:2018vyg,AtacamaCosmologyTelescope:2025blo}, and the cosmological evidence for dark matter (DM)~\cite{Cirelli:2024ssz,Planck:2018vyg,AtacamaCosmologyTelescope:2025blo}. Among the many proposed DM candidates, weakly interacting massive particles (WIMPs) remain a simple and predictive possibility, since their thermal production can naturally lead to an abundance of the order of the observed value~\cite{Arcadi:2024ukq,Cirelli:2024ssz}.

A viable realization of the WIMP paradigm is obtained by extending the SM by a single electroweak multiplet~\cite{Cirelli:2005uq,Cirelli:2009uv}. In such constructions, the phenomenology is largely fixed by the spin, weak isospin and hypercharge of the new field. The neutral component can be stabilized by an accidental symmetry, while electroweak radiative corrections lift the tree-level degeneracy among the components of the multiplet. This framework has been widely explored for scalar~\cite{Deshpande:1977rw,Barbieri:2006dq,LopezHonorez:2006gr,Hambye:2009pw} and fermionic~\cite{Cirelli:2005uq,Cirelli:2009uv,Bottaro:2021snn,Bottaro:2022one} candidates, whereas spin-1 realizations~\cite{Hambye:2008bq,Belyaev2019,Escalona:2024zkb} are relatively less studied because massive vector fields introduced through Proca terms are not protected by gauge invariance and may lead to perturbative unitarity violation at high energies. In effective field theory constructions, longitudinal vector scattering is kept unitary up to a high-energy cutoff by restricting non-minimal interactions and Higgs-portal couplings, establishing a controlled domain of validity for collider phenomenology.
The prototypical example is the minimal dark matter (MDM) construction, where a single electroweak multiplet can yield a stable neutral state with a radiatively generated compressed spectrum~\cite{Cirelli:2005uq,Cirelli:2009uv}. For a fermionic $SU(2)_L$ quintuplet with $Y=0$, the loop-induced splitting is typically of order $\mathcal{O}(100\,\mathrm{MeV})$, implying macroscopic lifetimes for the charged component and motivating disappearing-track searches at the LHC~\cite{Ostdiek:2015aga,Bharucha:2018lhy,Ibe:2012sx}.

A spin-1 dark matter candidate in the real five-dimensional representation of $SU(2)_L$ was introduced in Ref.~\cite{Escalona:2024zkb}. In that model, the neutral component of the quintuplet is stable due to an accidental symmetry, while the charged components become slightly heavier after radiative corrections. The model contains non-minimal interactions with the electroweak gauge fields and the Higgs sector, whose structure is constrained by perturbative unitarity. While thermal relic saturation pushes the preferred dark matter mass to tens of TeV, the sub-TeV and TeV domain remains relevant as a direct collider probe of electroweak vector multiplets and in multi-component dark matter scenarios.

The compressed spectrum of the vector quintuplet has direct collider consequences: because the mass splitting $\Delta m$ is at the sub-GeV scale, the decay width of the charged states is strongly suppressed, so that they propagate over macroscopic distances inside the detector before decaying into soft visible products. The exploration of long-lived particles (LLPs) has motivated a diverse program of dedicated experimental strategies across the LHC, ranging from central-detector signatures such as displaced vertices (DVs) and displaced tracks in both experimental searches and dedicated phenomenological recastings~\cite{ATLAS:2017nul,Cottin:2025sleptonsDVneutralino,Cottin:2022dvneutralino}, displaced leptons~\cite{CMS:2021rkn}, and heavy stable charged particles~\cite{Alimena:2019zri,Knapen2023}, to neutral LLP searches at proposed far-detector experiments including FASER~\cite{Feng:2017uoz}, MATHUSLA~\cite{Curtin:2018mvb}, CODEX-b~\cite{Gligorov:2017nwh}, and MoEDAL-MAPP~\cite{Pinfold:2019nqj,Staelens:2019rny}. 

In our scenario, a long-lived charged state decays into neutral dark matter with a sub-GeV mass splitting: the accompanying soft decay products fail standard vertex reconstruction, while the particle decays long before reaching external far detectors. Instead, the charged state traverses only the innermost tracking layers before decaying in flight, producing a distinctive \textit{disappearing-track} signature characterized by a short track segment that terminates abruptly without outer detector hits. 

This signature is well suited to inner-detector searches at the LHC~\cite{Ostdiek:2015aga,Bharucha:2018lhy,Ibe:2012sx,Belyaev:2020wok,Chiang:2020rcv,ATLAS:2022rme,ATLAS:2026hnb,CMS:2020atg}. While such searches have been extensively deployed by ATLAS and CMS to constrain long-lived charginos in supersymmetric scenarios, the same experimental strategy can be reinterpreted for other electroweak multiplets. Because the spin, electroweak representation, and production kinematics of the vector quintuplet differ substantially from those of supersymmetric charginos, a dedicated recasting is required.
In this work we study the disappearing-track phenomenology of the spin-1 quintuplet dark matter model. We focus on the singly charged component $V^\pm$, produced either directly or in the decay of the doubly charged state $V^{\pm\pm}$. Its decay into the neutral DM state $V^0$ and soft visible particles leads to the pixel-tracklet topology targeted by existing searches. We reinterpret the ATLAS disappearing-track search with the full Run-2 dataset~\cite{ATLAS:2026hnb}, originally designed for long-lived charginos, and derive the constraints from the Run-2 data, together with the projected sensitivity at higher luminosity, in the $(M_V,c\tau)$ and $(M_V,\Delta m)$ planes, where $M_V$ denotes the physical quintuplet mass parameter. 

The structure of the paper is as follows. In Sec.~\ref{sec:model}, we review the spin-1 quintuplet model and the origin of the mass splitting. In Sec.~\ref{sec:pheno}, we discuss the long-lived charged states and their decay modes. In Sec.~\ref{sec:methods}, we describe the simulation strategy and the reinterpretation of disappearing-track searches. In Sec.~\ref{sec:results}, we present the resulting constraints on the model parameters. We conclude in Sec.~\ref{sec:conc}.

\section{The model}
\label{sec:model}

In this section we summarize the ingredients of the spin-1 quintuplet model relevant for the disappearing-track analysis. This section follows Ref.~\cite{Escalona:2024zkb}, where the construction was introduced and studied in detail. The SM is supplemented by a massive vector multiplet $V_\mu$ transforming in the five-dimensional representation of $SU(2)_L$ with vanishing hypercharge,
\begin{equation}
    V_\mu \sim (\mathbf{5},0)
    \quad \text{under} \quad
    SU(2)_L \times U(1)_Y .
\end{equation}
We write its components as
\begin{equation}
    V_\mu =
    \left(
    V_\mu^{++},
    V_\mu^{+},
    V_\mu^{0},
    V_\mu^{-},
    V_\mu^{--}
    \right)^T ,
\end{equation}
where the superscript denotes the electric charge. The field $V_\mu$ is not a gauge boson. Rather, it is a massive vector field transforming homogeneously under $SU(2)_L$,
\begin{equation}
    V_\mu(x) \to V'_\mu(x)=U(x)V_\mu(x),
    \qquad U(x)\in SU(2)_L ,
\end{equation}
with the covariant derivative
\begin{equation}
    (D_\mu V_\nu)^i =
    \left(
    \partial_\mu \delta^{ij}
    +\frac{i g}{2}(T^a)^{ij}W^a_\mu
    \right)V^j_\nu ,
    \label{eq:covariant_derivative}
\end{equation}
where $i,j=1,\ldots,5$, and $T^a$ denote the generators of the five-dimensional representation of $SU(2)_L$, normalized such that the electric charge is $Q=T^3/2$, \ie $T^3 = \mathrm{diag}(4,2,0,-2,-4)$. This is the normalization encoded in our \textsc{FeynRules} implementation and in the resulting UFO model.

The most general effective Lagrangian for a massive vector multiplet transforming homogeneously under $SU(2)_L$ contains minimal gauge couplings, non-minimal gauge interactions, and general vector self-interactions:
\begin{align}
    \mathcal{L} &= \mathcal{L}_{\mathrm{SM}} 
    - \frac{1}{4} (V_{\mu\nu}^i)^\dagger V^{\mu\nu i} 
    + \frac{1}{2}m_V^2 (V_\mu^i)^\dagger V^{\mu i} 
    + \frac{1}{\xi}(D_\mu V^\mu)^\dagger(D_\nu V^\nu) 
    - i g \kappa W_{\mu\nu}^a (V^{\mu\dagger} T^a V^\nu) \nonumber \\[2mm]
    &\quad + \frac{\lambda_{HV}}{2}(\Phi^\dagger\Phi)(V_\mu^\dagger V^\mu) 
    + \alpha_1 (V_\mu^\dagger V^\mu)(V_\nu^\dagger V^\nu) 
    + \alpha_2 (V_\mu^\dagger V_\nu)(V^{\dagger\mu}V^\nu) 
    + \alpha_3 (V_\mu^\dagger V_\nu)(V^{\dagger\nu}V^\mu),
    \label{eq:full_lagrangian}
\end{align}
where $V_{\mu\nu}^i = (D_\mu V_\nu)^i - (D_\nu V_\mu)^i$, $W_{\mu\nu}^a$ is the $SU(2)_L$ field strength tensor, $\kappa$ is the non-minimal coupling to the gauge field, $1/\xi$ is the coefficient of the additional kinetic structure, $\lambda_{HV}$ is the Higgs-portal coupling, and $\alpha_{1,2,3}$ parameterize the quartic vector self-interactions.

Since $V_\mu$ is not protected by a gauge symmetry, amplitudes involving its longitudinal polarizations grow with energy. Preserving tree-level perturbative unitarity in the elastic ($V_L V_L \to V_L V_L$) and inelastic ($V_L V_L \to W_L W_L / Z_L Z_L / hh$) channels requires:
\begin{equation}
    \kappa = 0, \qquad \alpha_2 = 0, \qquad \alpha_1 = -\alpha_3 \equiv \alpha,
\end{equation}
along with the bounds
\begin{equation}
    |\alpha| \lesssim 4\pi, \qquad |\lambda_{HV}| \lesssim 4\pi, \qquad \frac{1}{\xi} \in [-10, -3/2].
\end{equation}
In this work we restrict ourselves to this unitarity-compatible regime. 
After electroweak symmetry breaking, the Higgs portal modifies the common tree-level mass of the quintuplet components according to
\begin{equation}
    M_V^2=m_V^2+\frac{1}{2}\lambda_{HV}v^2 .
    \label{eq:physical_mass}
\end{equation}
At tree level, all components of \(V_\mu\) are degenerate.

\begin{figure}[!htbp]
    \centering
    \includegraphics[width=0.75\linewidth]{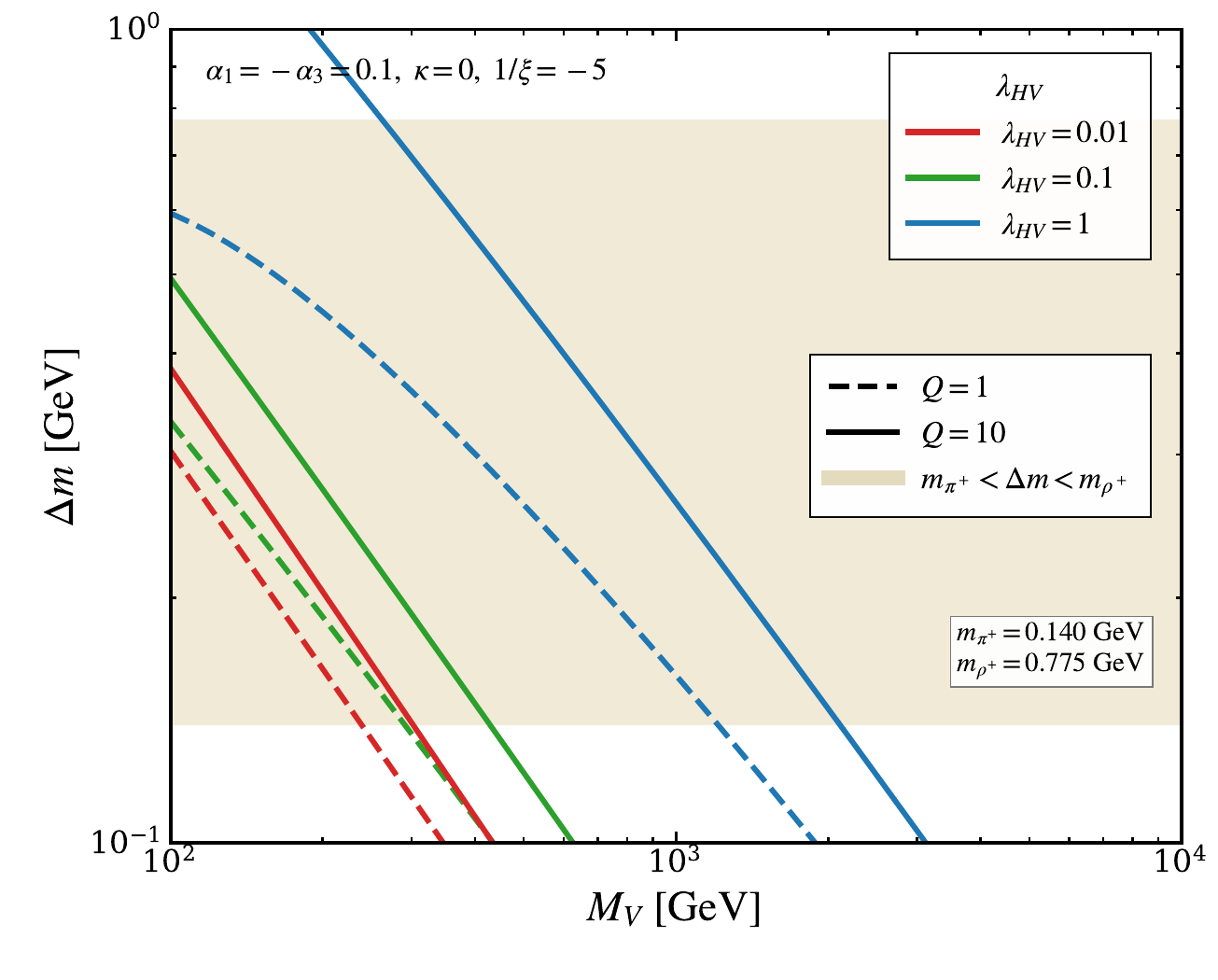}
\caption{One-loop radiative mass splitting $\Delta m = M_{V^\pm} - M_{V^0}$ for the singly charged vector quintuplet state as a function of the physical mass $M_V$, computed in dimensional regularization following Ref.~\cite{Escalona:2024zkb}. Curves correspond to Higgs portal couplings $\lambda_{HV} = 0.01$ (red), $\lambda_{HV} = 0.1$ (green), and $\lambda_{HV} = 1.0$ (blue), with vector interaction parameters fixed to the tree-level unitarity-compatible benchmark $\alpha_1 = -\alpha_3 = 0.1$, $\kappa = 0$, and $1/\xi = -5$. For each coupling, the dashed and solid curves denote renormalization scale choices $\mu = Q \cdot M_V$ with $Q = 1$ and $Q = 10$, respectively, illustrating the one-loop scale uncertainty. The horizontal shaded band delineates the intermediate mass-splitting window between the charged-pion threshold ($m_{\pi^+} \simeq 140~\mathrm{MeV}$) and the $\rho$-meson threshold ($m_{\rho^+} \simeq 770~\mathrm{MeV}$).}
    \label{fig:splitVmv}
\end{figure}

Radiative corrections lift the tree-level degeneracy between the neutral and charged components. We define the mass splitting relevant for the singly charged state as
\begin{equation}
    \Delta m \equiv M_{V^\pm}-M_{V^0}.
\end{equation}
The one-loop mass splitting is computed from the real, finite and transverse parts of the self-energies of the neutral and charged vector components. The self-energies were evaluated with \textsc{FeynArts} and \textsc{FeynCalc}~\cite{Hahn:2000kx,Shtabovenko:2020gxv}, using dimensional regularization. The splitting can be expressed as
\begin{equation}
\Delta m =
M_{V}
\sum_{n=1}^{\infty}
(-1)^{n}
\binom{1/2}{n}
\left[
\left(
\frac{\Sigma^{+}(M_{V}^{2})}{M_{V}^{2}}
\right)^{n}
-
\left(
\frac{\Sigma^{0}(M_{V}^{2})}{M_{V}^{2}}
\right)^{n}
\right],
\label{eq:split}
\end{equation} 
where $\binom{1/2}{n}$ is the generalized binomial coefficient and $\Sigma^{+,0}$ are the self-energies of the charged and neutral components.
The resulting splitting is typically at the sub-GeV scale over the collider-relevant mass range. For masses below the TeV scale, \(\Delta m\) is of the order of hundreds of MeV, while it decreases for larger values of \(M_V\), as shown in Fig.~\ref{fig:splitVmv}.

The neutral component \(V^0\) is the dark matter candidate of the model. The choice \(Y=0\) makes \(V^0\) electrically neutral and removes tree-level interactions with nucleons mediated by the \(Z\) boson, thus evading the strong direct-detection bounds on \(Z\) exchange. In addition, since the Lagrangian contains the new vector field only in pairs, an accidental \(Z_2\) symmetry is present. Under this symmetry the vector quintuplet is odd, while all SM fields are even. Consequently, the lightest component of the quintuplet is stable. \(V^0\) saturates the observed relic abundance only for masses of tens of TeV. In the sub-TeV and TeV region probed at the LHC it is therefore a subdominant component of the dark matter density.

This motivates collider searches as a complementary probe of the model. Since \(V^0\) has no tree-level coupling to the \(Z\) boson, its spin-independent scattering on nuclei proceeds through Higgs exchange, with a cross section proportional to \(\lambda_{HV}^2\) that lies below the limits of direct-detection experiments~\cite{XENON:2018voc,LZ:2022lsv}; in the mass range considered here the expected rate is further reduced by the small relic fraction. The same suppression applies, quadratically, to indirect-detection signals.

\section{LLP Phenomenology: Disappearing Charged Tracks}\label{sec:pheno}

As discussed in the previous section, radiative corrections generate a small mass splitting $\Delta m$ between the charged and neutral components of the quintuplet. This splitting controls which leptonic and hadronic decay channels are kinematically accessible and therefore largely determines the total width $\Gamma_{\text{tot}}$ of the charged states.

The multiplet also contains a doubly charged component $V^{\pm\pm}$. Its splitting with respect to $V^\pm$ is larger than $\Delta m$~\cite{Escalona:2024zkb}, so that it decays before reaching the pixel detector through $V^{\pm\pm} \to V^\pm \pi^\pm$. Every $V^{\pm\pm}$ therefore yields a $V^\pm$, which can be reconstructed as a disappearing track, and a pion too soft to affect the event selection. The channels $pp \to V^{\pm\pm} V^\mp$ and $pp \to V^{++} V^{--}$ thus contribute to the signal, and we include them together with $pp \to V^\pm V^0$ and $pp \to V^+ V^-$. The only splitting that enters our results is $\Delta m$, through the lifetime of $V^\pm$.

\subsection{Disappearing tracks}
\label{sec:pheno_dt}

Experimentally, the disappearing-track signature relies on reconstructing a short, isolated tracklet in the inner detector that terminates prior to the middle tracking subsystems~\cite{ATLAS:2017oal,ATLAS:2026hnb}. In the ATLAS Run-2 layout, the Pixel Detector consists of four cylindrical concentric barrel layers positioned at nominal radii $r \approx 33.25\text{ mm}$ (Insertable $B$-Layer), $50.5\text{ mm}$, $88.5\text{ mm}$, and $122.5\text{ mm}$, followed by the Semiconductor Tracker (SCT) whose innermost barrel layer begins at $r \approx 299\text{ mm}$~\cite{ATLAS:2017oal,ATLAS:2022rme}. In the four-layer signal region used in this work, the long-lived charged particle must traverse all four pixel layers to leave four space-point clusters, while requiring zero associated outer hits in the SCT to confirm its disappearance. The tracklet reconstruction efficiency is therefore a function of the transverse decay radius $r_{\text{decay}}$ of the charged particle: it is sizeable only for decays between the outermost pixel layer and the first SCT layer, $120\text{ mm} \lesssim r_{\text{decay}} \lesssim 300\text{ mm}$. In our recast this dependence is taken from the efficiency published by ATLAS (see Sec.~\ref{sec:tracklet_prob}).

Because the soft pion emitted in the compressed decay $V^\pm \to V^0 \pi^\pm$ cannot trigger the detector, the experimental signature relies on initial-state radiation (ISR) recoiling against the $V$ system. The hard recoil provides both the missing transverse momentum required for online event selection and a relativistic boost to the produced charged state. In the laboratory frame, the mean transverse decay length of the long-lived charged vector is governed by its relativistic transverse boost,
\begin{equation}
\ell_T = \beta_T \gamma \, c\tau = \frac{p_T}{M_V} \frac{\hbar c}{\Gamma_{\text{tot}}} = \frac{p_T}{M_V} \, c\tau ,
\label{eq:boosted_decay_length}
\end{equation}
where $p_T$ is the transverse momentum of $V^\pm$, $M_V$ is its physical mass, and $\Gamma_{\text{tot}}$ is the total decay width. 

The probability of reconstructing a disappearing track is therefore determined by the overlap between the exponential distribution of decay radii, with mean $\ell_T$, and this radial range. If the proper decay length is too short, the particles decay before reaching the fourth pixel layer; if it is too large, they reach the SCT and fail the disappearance requirement. The acceptance is largest when $\ell_T$ is comparable to the distance between the pixel detector and the SCT, which for the boosts of the selected events corresponds to proper decay lengths of order $0.1\text{--}1\text{ m}$.

\subsection{Decay modes of the charged particles}
Radiative corrections generate a mass splitting $\Delta m$ within the quintuplet, making $V^0$ the lightest state and rendering the charged components unstable. The neutral component is therefore stable, whereas the charged states decay to $V^0$ through an off-shell $W$ boson. For the benchmark scenarios considered here, the relevant decay channels of the singly charged state $V^{+}$ are:
\begin{itemize}
    \item the leptonic three-body mode $V^+ \to V^0\,\ell^+\,\nu_\ell$ (with $\ell=e,\mu$),
    \item hadronic modes, including the two-body channel $V^+\to V^0\pi^+$.
\end{itemize}

Since the values of $\Delta m$ of interest lie in the hadronic regime, a reliable estimate of the partial widths must incorporate QCD effects \cite{Belyaev:2020wok}. We therefore employ the standard effective interaction that describes the coupling of the $W$ boson to the charged pion,
\begin{equation}
    \mathcal{L}_{\mathrm{eff}} \;=\; g f_{\pi}\,\frac{V_{ud}}{2\sqrt{2}}\,W^+_{\mu}\,\partial^{\mu}\pi^- \;+\; \text{h.c.}\,,
    \label{eq:eff_pion_lag}
\end{equation}
where $V_{ud}$ is the relevant CKM element and $f_{\pi} \simeq 130\text{ MeV}$ is the pion decay constant. In the compressed mass regime where $\Delta m = M_{V^\pm} - M_{V^0}$ exceeds the charged-pion mass $m_\pi$, the total decay width of the singly charged state $V^\pm$ is dominated by the two-body hadronic channel $V^\pm \to V^0 \pi^\pm$, with leptonic three-body modes being strongly phase-space suppressed. Using this effective $W\text{--}\pi$ mixing framework, the analytic tree-level decay width for a spin-1 quintuplet decaying into a charged pion is given by \cite{Escalona:2024zkb}:
\begin{align}
\Gamma(V^\pm \to V^0 \pi^\pm) &= \frac{g^4 f_\pi^2 |V_{ud}|^2 \Delta m^2 (2M_{V^+} - \Delta m)^2 \sqrt{(\Delta m^2 - m_\pi^2)\left[(2M_{V^+} - \Delta m)^2 - m_\pi^2\right]}}{512\pi M_{V^+}^5 M_W^4 (M_{V^+} - \Delta m)^2} \nonumber \\
&\quad \times \left[ 2(M_{V^+} - \Delta m)^2 (5M_{V^+}^2 - m_\pi^2) + (M_{V^+}^2 - m_\pi^2)^2 + (M_{V^+} - \Delta m)^4 \right],
\label{eq:width_exact}
\end{align}
where $g$ is the $SU(2)_L$ gauge coupling and $M_W$ is the $W$-boson mass. 
In the heavy-multiplet limit $M_{V^+} \gg \Delta m$, Eq.~\eqref{eq:width_exact} smoothly reduces to the characteristic phase-space scaling $\Gamma(V^\pm \to V^0 \pi^\pm) \simeq \frac{6\,G_F^2 f_\pi^2 |V_{ud}|^2}{\pi} \Delta m ^3 \sqrt{1 - m_\pi^2/\Delta m^2}$ \cite{Ibe:2012sx,Chiang:2020rcv}. The leptonic modes $V^\pm \to V^0 \ell^\pm \nu$ lead to the same disappearing-track signature, but for $\Delta m > m_\pi$ they are a small correction to the width: $2.4\%$ of the two-body width at $\Delta m = 166\text{ MeV}$ and about $10\%$ at $\Delta m = 140\text{ MeV}$. In this work we restrict ourselves to $\Delta m > m_\pi$ and approximate $\Gamma_{\text{tot}}$ by the two-body width.
To describe the decay in the event generation, the charged pion was introduced as a physical pseudoscalar state and the effective interaction in Eq.~\eqref{eq:eff_pion_lag} was implemented in \textsc{FeynRules}~\cite{Alloul:2013bka}. The proper decay length is $c\tau = \hbar c / \Gamma$; its effect on the signal efficiency is treated analytically, as described in Sec.~\ref{sec:tracklet_prob}.

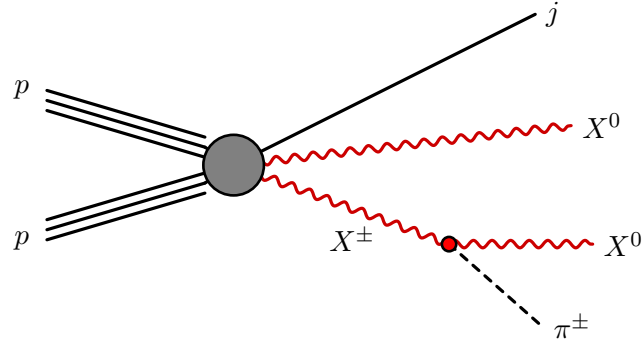
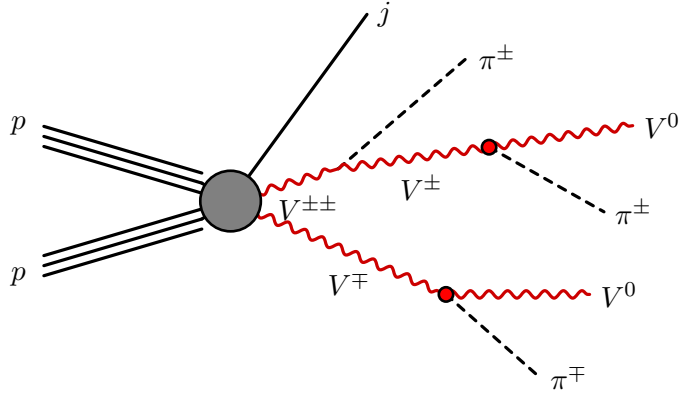
\begin{figure}[!htbp]
  \centering
  \begin{subfigure}{\linewidth}
    \centering
    \begin{tikzpicture}[thick, scale=0.95, line cap=round, line join=round, every node/.style={font=\small}]
      \coordinate (A) at (0,0);
      \coordinate (B) at (3,-1.1);

      \foreach \dy in {-0.14,0,0.14} {
        \draw[line width=1.2pt] (-2.6,0.9+\dy) -- (-0.4,0.25+\dy);
        \draw[line width=1.2pt] (-2.6,-0.9+\dy) -- (-0.4,-0.25+\dy);
      }
      \draw[line width=1.2pt] (-0.4,0.25) -- (A);
      \draw[line width=1.2pt] (-0.4,-0.25) -- (A);
      \node[anchor=east] at (-2.7,1.05) {$p$};
      \node[anchor=east] at (-2.7,-1.05) {$p$};

      \draw[line width=1.2pt] (A) -- (4.2,2.1);
      \node[anchor=west] at (4.2,2.1) {$j$};

      \draw[red!80!black,decorate,decoration={snake,amplitude=1.4pt,segment length=7pt},line width=1.2pt]
        (A) -- (4.7,0.55) node[anchor=west,black] {$X^{0}$};

      \draw[red!80!black,decorate,decoration={snake,amplitude=1.4pt,segment length=7pt},line width=1.2pt]
        (A) -- (B) node[pos=0.55,below=3pt,black] {$X^{\pm}$};

      \draw[red!80!black,decorate,decoration={snake,amplitude=1.4pt,segment length=7pt},line width=1.2pt]
        (B) -- (5.0,-1.1) node[anchor=west,black] {$X^{0}$};

      \draw[dashed,line width=1.2pt] (B) -- (4.3,-2.25) node[anchor=west] {$\pi^{\pm}$};

      \filldraw[fill=grey,draw=black,line width=1pt] (A) circle (12pt);
      \filldraw[fill=red,draw=black,line width=1pt] (B) circle (0.1);
    \end{tikzpicture}
    \caption{$pp\to X^{\pm}X^{0}j$ with $X^{\pm}\to X^{0}\pi^{\pm}$, for $X=\tilde{\chi}_1$ (wino benchmark) and $X=V$ (vector quintuplet).}
    \label{fig:feyncharg}
  \end{subfigure}

  \vspace{0.6em}

  \begin{subfigure}{\linewidth}
    \centering
    \begin{tikzpicture}[thick, scale=0.95, line cap=round, line join=round, every node/.style={font=\small}]
      \coordinate (A) at (0,0);
      \coordinate (C) at (1.5,0.45);
      \coordinate (B1) at (3.6,0.75);
      \coordinate (B2) at (3,-1.3);

      \foreach \dy in {-0.14,0,0.14} {
        \draw[line width=1.2pt] (-2.6,0.9+\dy) -- (-0.4,0.25+\dy);
        \draw[line width=1.2pt] (-2.6,-0.9+\dy) -- (-0.4,-0.25+\dy);
      }
      \draw[line width=1.2pt] (-0.4,0.25) -- (A);
      \draw[line width=1.2pt] (-0.4,-0.25) -- (A);
      \node[anchor=east] at (-2.7,1.05) {$p$};
      \node[anchor=east] at (-2.7,-1.05) {$p$};

      \draw[line width=1.2pt] (A) -- (1.9,2.6);
      \node[anchor=west] at (1.9,2.6) {$j$};

      \draw[red!80!black,decorate,decoration={snake,amplitude=1.4pt,segment length=7pt},line width=1.2pt]
        (A) -- (C) node[pos=0.72,below=3pt,black] {$V^{\pm\pm}$};
      \draw[dashed,line width=1.2pt] (C) -- (3.3,2.0) node[anchor=west] {$\pi^{\pm}$};
      \draw[red!80!black,decorate,decoration={snake,amplitude=1.4pt,segment length=7pt},line width=1.2pt]
        (C) -- (B1) node[pos=0.55,below=3pt,black] {$V^{\pm}$};
      \draw[red!80!black,decorate,decoration={snake,amplitude=1.4pt,segment length=7pt},line width=1.2pt]
        (B1) -- (5.6,1.05) node[anchor=west,black] {$V^{0}$};
      \draw[dashed,line width=1.2pt] (B1) -- (5.2,-0.15) node[anchor=west] {$\pi^{\pm}$};

      \draw[red!80!black,decorate,decoration={snake,amplitude=1.4pt,segment length=7pt},line width=1.2pt]
        (A) -- (B2) node[pos=0.55,below=3pt,black] {$V^{\mp}$};
      \draw[red!80!black,decorate,decoration={snake,amplitude=1.4pt,segment length=7pt},line width=1.2pt]
        (B2) -- (5.0,-1.3) node[anchor=west,black] {$V^{0}$};
      \draw[dashed,line width=1.2pt] (B2) -- (4.3,-2.45) node[anchor=west] {$\pi^{\mp}$};

      \filldraw[fill=grey,draw=black,line width=1pt] (A) circle (12pt);
      \filldraw[fill=red,draw=black,line width=1pt] (B1) circle (0.1);
      \filldraw[fill=red,draw=black,line width=1pt] (B2) circle (0.1);
    \end{tikzpicture}
    \caption{$pp\to V^{\pm\pm}V^{\mp}j$, where the doubly charged state decays promptly, $V^{\pm\pm}\to V^{\pm}\pi^{\pm}$.}
    \label{fig:feynman-tikz}
  \end{subfigure}

  \caption{Signal topologies considered in this work. (a) Disappearing-track topology common to the wino benchmark used for validation and to the vector quintuplet. (b) Production involving a doubly charged state, specific to the quintuplet. Red dots mark the displaced decay of the long-lived charged state; dashed lines denote the soft pions.}
  \label{fig:feynman-combined}
\end{figure}

\section{Methodology}\label{sec:methods}

We reinterpret the ATLAS search for long-lived charginos in final states with a disappearing track, based on the full Run-2 dataset of $137~\mathrm{fb}^{-1}$~\cite{ATLAS:2026hnb}, following the general strategy of Refs.~\cite{Belyaev:2020wok,Chiang:2020rcv}. In the anomaly-mediated supersymmetry breaking (AMSB) interpretation the signal is the electroweak production of nearly mass-degenerate $\tilde{\chi}_1^{\pm}$ and $\tilde{\chi}_1^{0}$ in association with initial-state radiation (ISR). In our model the analogous processes are
\[
pp \to V^{\pm}V^{0}, \quad V^{+}V^{-}, \quad V^{\pm\pm}V^{\mp}, \quad V^{++}V^{--} \quad (+\,\text{jets}),
\]
which yield the same signature: a high-$p_T$ jet, large missing transverse momentum, and a short track from the charged state (Fig.~\ref{fig:feynman-combined}). The selection is therefore the same for both models, while all efficiencies are recomputed for the vector quintuplet to account for the differences in spin, production kinematics and boost.

\subsection{Event generation}
\label{sec:generation}
Events are generated at leading order with \textsc{MadGraph5\_aMC@NLO}~v2.9.21~\cite{Alwall:2011uj,Alwall:2014hca} at $\sqrt{s} = 13\text{ TeV}$ with the NNPDF2.3 LO parton distribution functions~\cite{Ball:2012cx}, including up to two additional partons in the matrix element. They are showered and hadronized with \textsc{Pythia~8.3}~\cite{Bierlich:2022pfr}, and the jet multiplicities are merged with the MLM prescription~\cite{Alwall:2007fs,Mangano:2006rw} using a merging scale of one quarter of the mass of the produced states.

The quintuplet is simulated with the UFO model obtained with \textsc{FeynRules}~\cite{Alloul:2013bka} from Eq.~\eqref{eq:full_lagrangian}, supplemented by the effective pion interaction of Eq.~\eqref{eq:eff_pion_lag}, with $\kappa = 0$, as required by perturbative unitarity (Sec.~\ref{sec:model}), and masses $M_V$ between $100$ and $1800~\mathrm{GeV}$ in steps of $100~\mathrm{GeV}$. At this order the production cross section is fixed by the gauge couplings: the $1/\xi$ term does not contribute for on-shell vectors, and the Higgs-mediated contribution proportional to $\lambda_{HV}$ vanishes for massless quarks. We verified numerically that the cross section is independent of both parameters. The signal is normalized to the leading-order cross section after merging. For the validation we use a pure-wino benchmark with a mass splitting of $160~\mathrm{MeV}$ and chargino masses between $100$ and $1100~\mathrm{GeV}$, normalized to the NLO+NLL cross sections~\cite{Fuks:2012qx,Fuks:2013vua}, as in the ATLAS interpretation.

A single sample is generated for each mass, with $\Delta m = 166~\mathrm{MeV}$, a reference value taken from the fermionic quintuplet~\cite{Cirelli:2005uq}, and $M_{V^{\pm\pm}} - M_{V^\pm} = 3\Delta m \simeq 0.5~\mathrm{GeV}$, for which the proper decay length of the doubly charged state is $0.6~\mathrm{mm}$. The lifetime of the charged state is not scanned at generation level: its effect on the efficiency is computed analytically from the same events, as described in Sec.~\ref{sec:tracklet_prob}. This is justified because the mass splitting affects the selection only through the lifetime; the kinematics of the heavy states are insensitive to a sub-GeV change in $\Delta m$.

\subsection{Detector emulation and event selection}
\label{sec:selection}
The detector response is emulated with an adaptation of the \textsc{ToyDetector} framework~\cite{TOYDETECTOR} to the ATLAS geometry. Stable visible particles with $|\eta| < 4.5$ are used to build the reconstructed objects. Electrons and muons are required to have $p_T > 10~\mathrm{GeV}$, $|\eta| < 2.5$ and to be isolated from charged activity within $\Delta R = 0.2$. The remaining particles are clustered into jets with \textsc{FastJet}~\cite{Cacciari:2011ma} using the anti-$k_T$ algorithm~\cite{Cacciari:2008gp} with $R = 0.4$, keeping jets with $p_T > 20~\mathrm{GeV}$. The missing transverse momentum $\vec{p}_T^{\mathrm{miss}}$ is the negative vector sum of the transverse momenta of the jets and isolated leptons. The neutral states $\tilde{\chi}_1^0$ and $V^0$ are invisible, and the long-lived charged state is excluded from the jets and from $\vec{p}_T^{\mathrm{miss}}$, as is the tracklet in the ATLAS analysis.

We use the signal region $\mathrm{SR}^{4\mathrm{High}}$ of the search, which requires a four-layer tracklet and places no requirement on the soft pion; it is the region with the lowest background and drives the sensitivity for the lifetimes considered here. The region with intermediate missing transverse momentum, $\mathrm{SR}^{4\mathrm{Mid}}$, was also evaluated and does not improve the exclusion. The three-layer regions, which partly rely on the reconstruction of the pion, are not included. The selection is summarized in Table~\ref{tab:selection}. A tracklet candidate is a long-lived charged state passing the kinematic and isolation requirements listed there, where the isolation sums exclude its own decay products. States $V^\pm$ produced in the decay of a $V^{\pm\pm}$ are treated in the same way as those produced directly.

\begin{table}[!htbp]
\centering
\setlength{\tabcolsep}{8pt}
\begin{tabular}{lc}
\hline\hline
 & $\mathrm{SR}^{4\mathrm{High}}$ \\
\hline
Isolated leptons ($p_T > 10$ GeV) & 0 \\
Leading-jet $p_T$ ($|\eta| < 2.8$) & $> 100$ GeV \\
$\Delta\phi(j_1, \vec{p}_T^{\mathrm{miss}})$ & $> 1.0$ \\
$p_T^{\mathrm{miss}}$ & $> 300$ GeV \\
\hline
Tracklet $p_T$ & $> 60$ GeV \\
Tracklet $|\eta|$ & $0.1$--$1.9$ \\
$\Delta R(\text{tracklet}, \text{jet})$, jets with $p_T > 20$ GeV & $> 0.4$ \\
Track isolation ($\Delta R < 0.4$) & $\sum p_T^{\text{ch}} < 0.04\, p_T$ \\
Calorimeter isolation ($\Delta R < 0.4$) & $\sum E_T < 5$ GeV \\
\hline
Expected background & $0.68 \pm 0.14$ \\
Observed events & 3 \\
$S^{95}_{\rm obs}$ ($S^{95}_{\rm exp}$) & 6.0 (3.7) \\
\hline\hline
\end{tabular}
\caption{Event and tracklet selection of the signal region $\mathrm{SR}^{4\mathrm{High}}$ used in this work, together with the background expectation, the observed yield, and the model-independent 95\% C.L. upper limit on the number of signal events reported by ATLAS~\cite{ATLAS:2026hnb}.}
\label{tab:selection}
\end{table}

\subsection{Tracklet reconstruction probability}
\label{sec:tracklet_prob}
For a given proper decay length $c\tau$, the transverse decay radius of a candidate follows an exponential distribution with mean $\ell_T = (p_T/M)\,c\tau$, Eq.~\eqref{eq:boosted_decay_length}. The probability that it is reconstructed as a four-layer tracklet is obtained by integrating this distribution over the tracklet reconstruction efficiency measured by ATLAS as a function of the decay radius (Fig.~2(a) of Ref.~\cite{ATLAS:2026hnb}). Writing the efficiency as $\varepsilon_k$ in the radial bin $[r_k^{\min}, r_k^{\max}]$,
\begin{equation}
    P_{\rm trk}(c\tau) = \sum_k \varepsilon_k \left[ \exp\!\left(-\frac{r_k^{\min}}{\ell_T}\right) - \exp\!\left(-\frac{r_k^{\max}}{\ell_T}\right) \right].
    \label{eq:ptrk}
\end{equation}
The efficiency vanishes below $r \simeq 120~\mathrm{mm}$, lies between $0.5$ and $0.8$ up to $r \simeq 290~\mathrm{mm}$, and is applied uniformly over the $|\eta|$ acceptance. An event passing the selection is assigned the probability that at least one of its candidates is reconstructed,
\begin{equation}
    w(c\tau) = 1 - \prod_i \left[ 1 - P_{{\rm trk},i}(c\tau) \right],
    \label{eq:event_weight}
\end{equation}
and the total efficiency is $\epsilon(M, c\tau) = \sum w(c\tau) / N_{\rm gen}$, where $N_{\rm gen}$ is the number of events after merging. It is evaluated for 61 values of $c\tau$ between $1~\mathrm{mm}$ and $10~\mathrm{m}$ using the same events.

\subsection{Signal yield and limits}
\label{sec:stat}
The expected number of signal events is
\begin{equation}
  N_{\mathrm{sig}}(M, c\tau) = \sigma(M) \times \epsilon(M, c\tau) \times \mathcal{L},
  \label{eq:nsig}
\end{equation}
where $\sigma$ is the total production cross section and $\mathcal{L}$ the integrated luminosity. A point is excluded at 95\% C.L. when $N_{\mathrm{sig}}$ exceeds the ATLAS upper limit $S^{95}_{\rm obs}$ of Table~\ref{tab:selection}; the expected exclusion is obtained with $S^{95}_{\rm exp}$. For each lifetime, the mass limit is obtained by interpolation between the simulated masses. Limits in the $(M_V, \Delta m)$ plane follow from $c\tau = \hbar c/\Gamma(M_V, \Delta m)$ with the width of Eq.~\eqref{eq:width_exact}. Theoretical uncertainties on the leading-order cross section (scale and parton distribution functions) are not included.

\subsection{Validation}
\label{sec:validation}
\begin{figure}[!htbp]
    \centering
    \includegraphics[width=0.8\linewidth]{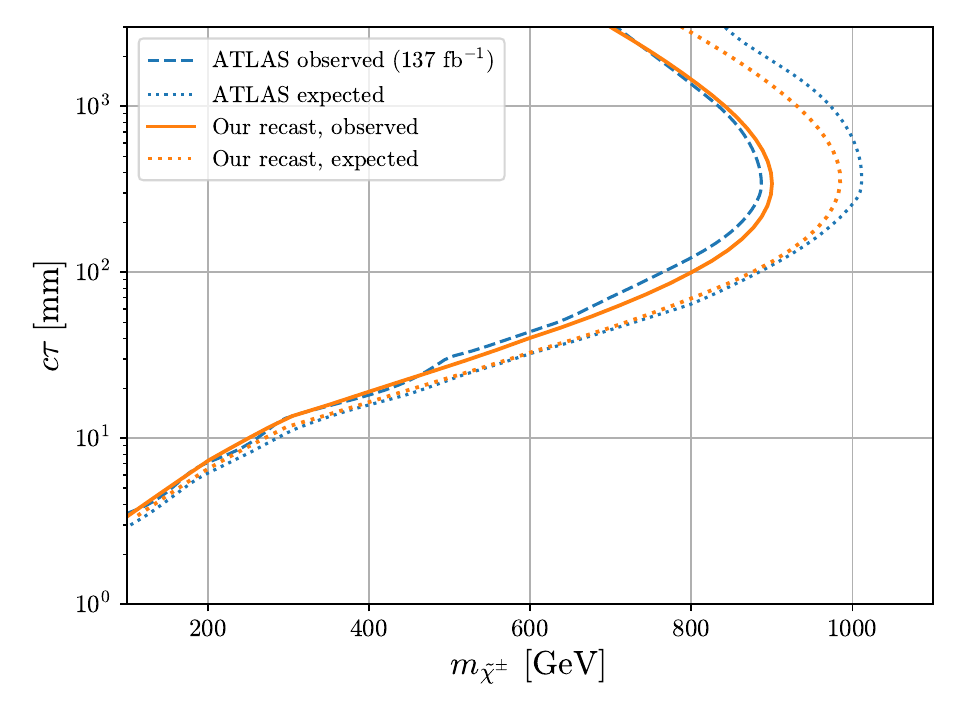}
    \caption{Validation of the recast for pure winos: 95\% C.L. exclusion in the plane of the chargino mass and proper decay length at $\sqrt{s}=13~\mathrm{TeV}$ and $\mathcal{L} = 137~\mathrm{fb}^{-1}$. The observed (solid) and expected (dotted) limits obtained in this work (orange) are compared with the observed (dashed) and expected (dotted) ATLAS limits (blue) from Ref.~\cite{ATLAS:2026hnb}. A proper decay length of $300~\mathrm{mm}$ corresponds to a lifetime of $1~\mathrm{ns}$.}
    \label{fig:recast_chargino}
\end{figure}

Figure~\ref{fig:recast_chargino} compares the wino exclusion obtained with this procedure with the ATLAS result. For $c\tau = 60~\mathrm{mm}$ ($\tau = 0.2~\mathrm{ns}$), the value predicted for pure winos, we exclude chargino masses up to $700~\mathrm{GeV}$ ($765~\mathrm{GeV}$ expected), to be compared with $670~\mathrm{GeV}$ ($770~\mathrm{GeV}$) from ATLAS. Around $c\tau = 300~\mathrm{mm}$ ($\tau = 1~\mathrm{ns}$), where the sensitivity is maximal, we obtain $900~\mathrm{GeV}$ ($985~\mathrm{GeV}$), against $880~\mathrm{GeV}$ ($1020~\mathrm{GeV}$). Over the range $5~\mathrm{mm} < c\tau < 2~\mathrm{m}$ the observed and expected mass limits agree with those of ATLAS within $6\%$. Exact agreement is not expected, since our limits come from a single-bin counting experiment whereas ATLAS performs a combined fit to the tracklet $p_T$ spectrum in all signal regions.

\section{Results}
\label{sec:results}

\begin{figure}[!htbp]
    \centering
    \includegraphics[width=0.72\linewidth]{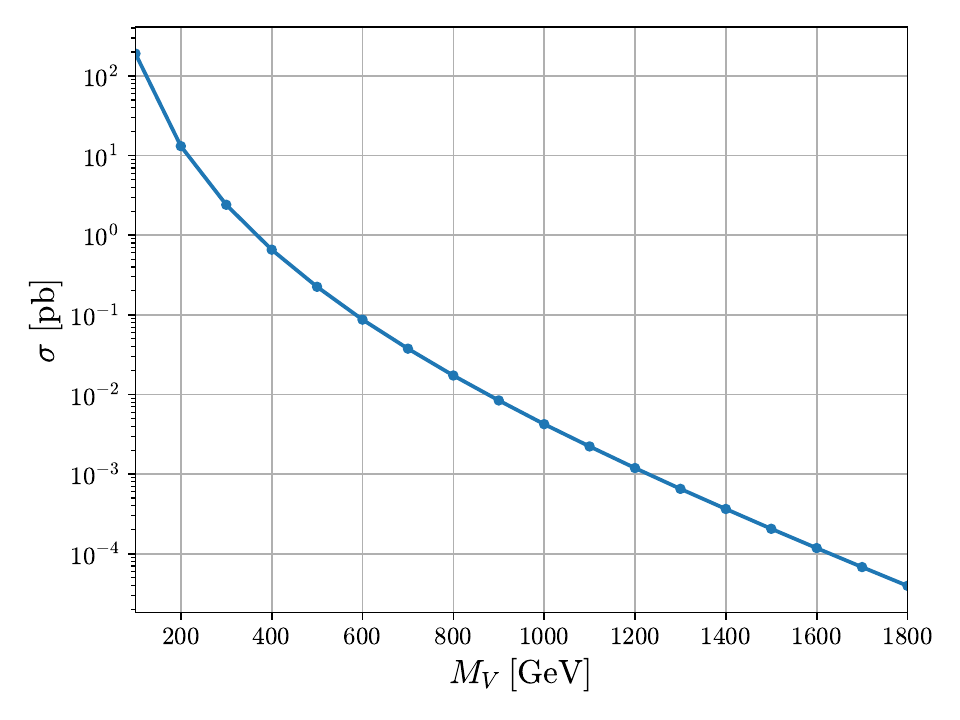}
    \caption{Total leading-order pair-production cross section of the vector quintuplet states as a function of the physical mass $M_V$ at $\sqrt{s} = 13\text{ TeV}$, for $\kappa = 0$, after merging the samples with up to two additional partons.}
    \label{fig:xsection_vs_mass}
\end{figure}

We now apply the analysis of Sec.~\ref{sec:methods} to the vector quintuplet. The leading-order production cross section, summed over all pair-production modes, is shown in Fig.~\ref{fig:xsection_vs_mass}; it ranges from $190~\mathrm{pb}$ at $M_V = 100~\mathrm{GeV}$ to $2.4~\mathrm{pb}$ at $300~\mathrm{GeV}$, $4.2~\mathrm{fb}$ at $1~\mathrm{TeV}$ and $0.04~\mathrm{fb}$ at $1.8~\mathrm{TeV}$.

We present two sets of results. The first is the constraint from the Run-2 data, obtained by comparing $N_{\mathrm{sig}}$ at $\mathcal{L} = 137~\mathrm{fb}^{-1}$ with the ATLAS limits of Table~\ref{tab:selection}. The second is a projection to $\mathcal{L} = 3~\mathrm{ab}^{-1}$, for which we keep $\sqrt{s} = 13~\mathrm{TeV}$ and the Run-2 selection and tracklet efficiency. We do not move to $\sqrt{s} = 14~\mathrm{TeV}$ because the tracklet efficiency and the background are only known for the Run-2 conditions; since the production cross section grows with the collision energy, this choice is conservative for the signal rate. The projection should therefore be read as the reach of a Run-2-like detector and analysis with a larger dataset.

The projection requires an assumption on the background, which in this search is purely instrumental. In $\mathrm{SR}^{4\mathrm{High}}$ it consists, in similar proportions, of fake tracklets from random combinations of pixel hits and of muons that are not reconstructed in the outer detectors~\cite{ATLAS:2026hnb}. Both are determined from data and cannot be reproduced in a fast simulation~\cite{Chiang:2020rcv}. We therefore consider two benchmarks. As an idealized ceiling we take a vanishing background, for which the 95\% C.L. limit corresponds to $N_{\mathrm{sig}} \ge 3$~\cite{Chiang:2020rcv,Belyaev:2020wok}. As a second benchmark we scale the $\mathrm{SR}^{4\mathrm{High}}$ background linearly with the luminosity, $B \approx 15$; assuming that the observed count equals the background and neglecting its uncertainty, the $CL_s$ prescription~\cite{Read:2002hq} gives $N_{\mathrm{sig}} \ge 9.5$. Since the fake-tracklet rate is expected to grow with pileup, this benchmark is indicative rather than conservative.

\subsection{Constraints in the $c\tau$--$M_V$ plane}
\label{sec:results_ctau}

\begin{figure}[!htbp]
    \centering
    \includegraphics[width=0.72\linewidth]{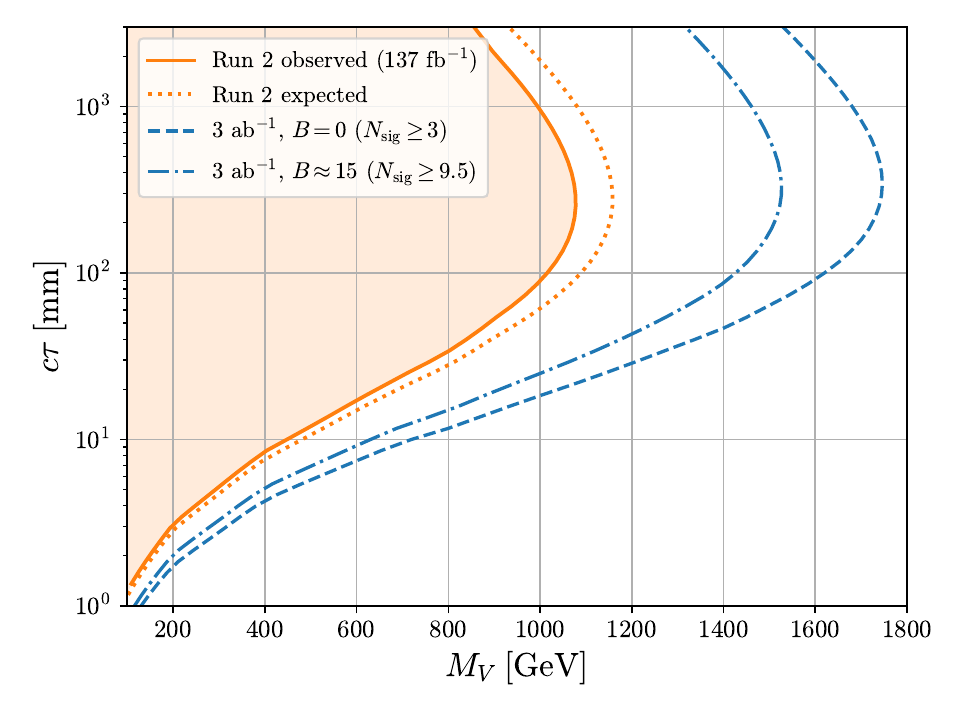}
    \caption{95\% C.L. limits on the vector quintuplet in the plane of the mass $M_V$ and the proper decay length $c\tau$ of $V^\pm$. The orange contours are the observed (solid) and expected (dotted) limits with $137~\mathrm{fb}^{-1}$; the shaded region is excluded. The blue contours are the projections for $3~\mathrm{ab}^{-1}$ without background ($N_{\mathrm{sig}} \ge 3$, dashed) and with background ($B \approx 15$, $N_{\mathrm{sig}} \ge 9.5$, dash-dotted).}
    \label{fig:ctau_mv_limit}
\end{figure}

Figure~\ref{fig:ctau_mv_limit} shows the constraints in the plane of the proper decay length of $V^\pm$ and the quintuplet mass. In this plane the result does not rely on the relation between lifetime and mass splitting, and holds for any $\Delta m$ small enough for the decay products of $V^\pm$ not to affect the selection.

The Run-2 data exclude masses up to $M_V \simeq 1080~\mathrm{GeV}$ ($1160~\mathrm{GeV}$ expected), reached for $c\tau \simeq 250~\mathrm{mm}$, and the limit stays above $1~\mathrm{TeV}$ for $0.1~\mathrm{m} \lesssim c\tau \lesssim 1~\mathrm{m}$. For shorter lifetimes it is $890~\mathrm{GeV}$ at $c\tau = 50~\mathrm{mm}$, $450~\mathrm{GeV}$ ($480~\mathrm{GeV}$ expected) at $c\tau = 10~\mathrm{mm}$ and $200~\mathrm{GeV}$ at $c\tau = 3~\mathrm{mm}$; for longer lifetimes it is still $855~\mathrm{GeV}$ at $c\tau = 3~\mathrm{m}$. These limits are stronger than those for winos at the same lifetime, $930~\mathrm{GeV}$ compared with $700~\mathrm{GeV}$ at $c\tau = 60~\mathrm{mm}$, owing to the larger production cross section, even though the quintuplet is normalized at leading order and the winos at NLO+NLL. With $3~\mathrm{ab}^{-1}$ the maximum reach extends to $1750~\mathrm{GeV}$ for vanishing background and to $1530~\mathrm{GeV}$ for the extrapolated background, and at $c\tau = 10~\mathrm{mm}$ to $720$ and $630~\mathrm{GeV}$, respectively.

The shape of the boundary reflects how the laboratory decay length $\ell_T = (p_T/M_V)\,c\tau$ maps onto the detector geometry. For $c\tau \lesssim 10~\mathrm{mm}$ most $V^\pm$ decay before the fourth pixel layer, and only the boosted tail of the $p_T$ spectrum is reconstructed, which restricts the sensitivity to low masses. The reach is largest for $c\tau \sim 0.1\text{--}1~\mathrm{m}$, where $\ell_T$ matches the distance between the pixel detector and the SCT. For longer lifetimes an increasing fraction of the charged states reaches the SCT, and the efficiency decreases slowly, approximately as $1/c\tau$.

\subsection{Constraints in the $\Delta m$--$M_V$ plane}

\begin{figure}[!htbp]
    \centering
    \includegraphics[width=0.72\linewidth]{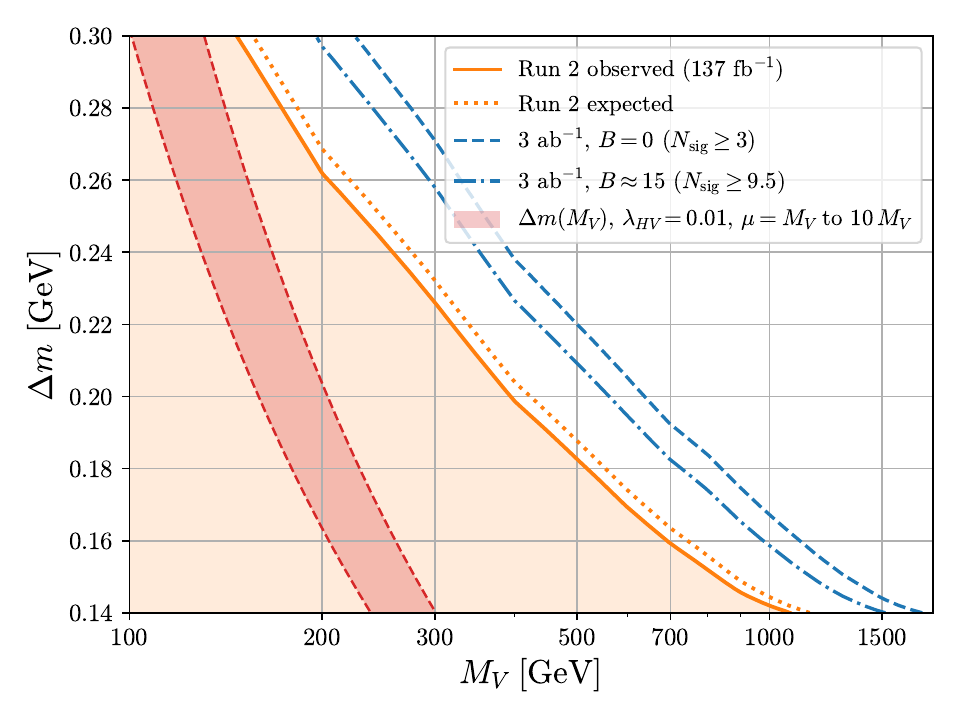}
    \caption{95\% C.L. limits on the vector quintuplet in the plane of the mass $M_V$ and the mass splitting $\Delta m$. The orange contours are the observed (solid) and expected (dotted) limits with $137~\mathrm{fb}^{-1}$; the shaded region is excluded. The blue contours are the projections for $3~\mathrm{ab}^{-1}$ without background ($N_{\mathrm{sig}} \ge 3$, dashed) and with background ($B \approx 15$, $N_{\mathrm{sig}} \ge 9.5$, dash-dotted). The red band is the one-loop radiative mass splitting of the model~\cite{Escalona:2024zkb} for $\lambda_{HV} = 0.01$, between the renormalization scales $\mu = M_V$ (lower edge) and $\mu = 10\,M_V$ (upper edge).}
    \label{fig:dm_mv_limit}
\end{figure}

The limits are translated into the $(M_V, \Delta m)$ plane through the width of Eq.~\eqref{eq:width_exact}, for mass splittings above the charged-pion threshold, $\Delta m \in [0.14, 0.30]~\text{GeV}$, and are shown in Fig.~\ref{fig:dm_mv_limit}. At fixed $M_V$, increasing $\Delta m$ enhances the width as $\Delta m^3 \sqrt{1 - m_\pi^2/\Delta m^2}$ and shortens the lifetime, until the decays occur before the outer pixel layer; this defines the upper edge of the excluded region. The proper decay length is $c\tau \simeq 220~\mathrm{mm}$ for $\Delta m = 140~\mathrm{MeV}$, $19~\mathrm{mm}$ for $166~\mathrm{MeV}$, $8~\mathrm{mm}$ for $200~\mathrm{MeV}$ and $2~\mathrm{mm}$ for $300~\mathrm{MeV}$, almost independently of $M_V$.

The Run-2 data exclude $M_V < 1075~\mathrm{GeV}$ ($1155~\mathrm{GeV}$ expected) for $\Delta m = 140~\mathrm{MeV}$, $M_V < 630~\mathrm{GeV}$ ($675~\mathrm{GeV}$) for $\Delta m = 166~\mathrm{MeV}$, $M_V < 395~\mathrm{GeV}$ for $\Delta m = 200~\mathrm{MeV}$ and $M_V < 205~\mathrm{GeV}$ for $\Delta m = 260~\mathrm{MeV}$. With $3~\mathrm{ab}^{-1}$ the limit at $\Delta m = 166~\mathrm{MeV}$ would reach $1020~\mathrm{GeV}$ for vanishing background and $890~\mathrm{GeV}$ for the extrapolated background.

Figure~\ref{fig:dm_mv_limit} also shows the one-loop radiative splitting $\Delta m(M_V)$ of the model (Sec.~\ref{sec:model}) for $\lambda_{HV} = 0.01$. Since the one-loop result depends sizeably on the renormalization scale (Fig.~\ref{fig:splitVmv}), the prediction is shown as a band between $\mu = M_V$ and $\mu = 10\,M_V$. The splitting decreases from $0.30~\mathrm{GeV}$ at $M_V \simeq 100\text{--}130~\mathrm{GeV}$ to the pion mass at $M_V \simeq 240\text{--}300~\mathrm{GeV}$, where the ranges correspond to the two scales, and in this range of splittings the whole band lies inside the region excluded by the Run-2 data. The model prediction is therefore excluded up to $M_V \simeq 240~\mathrm{GeV}$ for $\mu = M_V$, and for $120 \lesssim M_V \lesssim 300~\mathrm{GeV}$ for $\mu = 10\,M_V$; for this scale the splitting below $120~\mathrm{GeV}$ exceeds $0.32~\mathrm{GeV}$ and $V^\pm$ decays before the outer pixel layer. Larger values of $\lambda_{HV}$ shift the band towards larger splittings (Fig.~\ref{fig:splitVmv}): for $\lambda_{HV} = 0.1$ and $\mu = 10\,M_V$ the excluded interval is $215 \lesssim M_V \lesssim 430~\mathrm{GeV}$, while for $\lambda_{HV} = 1$ the splitting remains above $0.24~\mathrm{GeV}$ up to $M_V \simeq 1.1~\mathrm{TeV}$ and the prediction is not constrained. At the upper end of each interval the predicted splitting falls below $m_\pi$. There the hadronic mode is closed, $V^\pm$ decays through $V^\pm \to V^0 \ell^\pm \nu$ and its proper decay length is of the order of metres; this regime is outside the scope of this work, which is restricted to $\Delta m > m_\pi$.

\section{Conclusion}\label{sec:conc}

We have studied the disappearing-track signature of a spin-1 dark matter candidate in the real five-dimensional representation of $SU(2)_L$ with $Y=0$. Radiative corrections split the multiplet by an amount of the order of the pion mass, so that the singly charged state $V^\pm$ decays through $V^\pm \to V^0\pi^\pm$ after travelling a macroscopic distance in the inner detector.

We simulated the model with \textsc{MadGraph5\_aMC@NLO} and \textsc{Pythia}~8 and recast the four-layer signal region $\mathrm{SR}^{4\mathrm{High}}$ of the ATLAS full Run-2 search, including all pair-production modes of the charged states. The tracklet reconstruction probability is computed analytically from the published efficiency as a function of the decay radius, so that a single sample per mass covers the full range of lifetimes. The procedure reproduces the ATLAS limits for pure winos within $6\%$ in mass.

The Run-2 data exclude quintuplet masses up to $1080~\mathrm{GeV}$ for proper decay lengths around $250~\mathrm{mm}$, and up to $450~\mathrm{GeV}$ for $c\tau = 10~\mathrm{mm}$. In terms of the mass splitting, masses below $630~\mathrm{GeV}$ are excluded for $\Delta m = 166~\mathrm{MeV}$ and below $395~\mathrm{GeV}$ for $\Delta m = 200~\mathrm{MeV}$. The one-loop mass splitting predicted by the model for $\lambda_{HV} = 0.01$ is excluded for masses between $120~\mathrm{GeV}$ and $240\text{--}300~\mathrm{GeV}$, depending on the renormalization scale; above this range the two-body pion decay is closed. With $3~\mathrm{ab}^{-1}$ and a Run-2-like detector and analysis, the maximum reach would extend to $1750~\mathrm{GeV}$ for vanishing background and to $1530~\mathrm{GeV}$ for a linear extrapolation of the Run-2 background.

These results rely on a single-bin counting experiment with a simplified detector emulation and a leading-order signal normalization, and the projection assumes the Run-2 detector geometry. The analysis is restricted to $\Delta m > m_\pi$, where the leptonic modes can be neglected; for larger masses the predicted splitting falls below the pion mass and the charged state decays leptonically with a proper decay length of metres, a regime that deserves a dedicated study. A dedicated experimental analysis with the upgraded Inner Tracker and data-driven backgrounds will be needed to establish the definitive coverage of the vector quintuplet parameter space.

\section*{Acknowledgments}
We thank Giovanna Cottin for fruitful discussions on this work.
F.~H. is supported by ANID Beca de Doctorado Nacional grant No.~21250603. 
P.~E. is supported by CNPq grant No.~151612/2024-2. and the Fundaç\~ao de Amparo e Promoç\~ao da Ciência, Tecnologia e Inovaç\~ao do Rio Grande do Norte (FAPERN) via
the call number 06/2026 - FAPERN/IIP, with process number 10910022.000493/2024-04.
A.~Z. is supported by ANID-Chile FONDECYT grant No.~1230110 and ANID--Millennium Science Initiative Program--ICN2019\_044 (SAPHIR).
The authors used Claude (Anthropic, Claude Opus 5.5) to assist in writing parts of the analysis code, in cross-checking numerical results, and in drafting and editing parts of the text. The simulations were run by the authors, who directed the analysis, verified all results and revised the manuscript, and take full responsibility for its content.

\appendix

\bibliography{references}
\end{document}